\documentclass[%
 notitlepage,
 reprint,
 superscriptaddress,
 amsmath,amssymb,
 aps,
]{revtex4-2}

\usepackage{lipsum}
\usepackage{graphicx}
\usepackage{dcolumn}
\usepackage{bm}

\usepackage{epstopdf}
\usepackage{xcolor}

\newcommand{\bra}[1]{\langle #1 \vert}
\newcommand{\ket}[1]{\vert #1 \rangle}

\begin{document}

\preprint{APS/123-QED}

\title{Flat-band quantum communication induced by disorder}

\author{G. M. A. Almeida}
\email{gmaalmeida@fis.ufal.br}
\affiliation{%
 Instituto de F\'{i}sica, Universidade Federal de Alagoas, 57072-900 Macei\'{o}, AL, Brazil
}%

\author{R. F. Dutra}
\affiliation{%
 Instituto de F\'{i}sica, Universidade Federal de Alagoas, 57072-900 Macei\'{o}, AL, Brazil
}%

\author{A. M. C. Souza}
\affiliation{%
 Departamento de F\'{i}sica, Universidade Federal de Sergipe, 49100-000 S\~{a}o Crist\'{o}v\~{a}o, SE, Brazil
}%

\author{M. L. Lyra}
\affiliation{%
 Instituto de F\'{i}sica, Universidade Federal de Alagoas, 57072-900 Macei\'{o}, AL, Brazil
}%

\author{F. A. B. F. de Moura}
\affiliation{%
 Instituto de F\'{i}sica, Universidade Federal de Alagoas, 57072-900 Macei\'{o}, AL, Brazil
}%

\begin{abstract}

We show that a qubit transfer protocol can be realized through a flat band hosted by a disordered $XX$ spin-1/2 diamond chain. In the absence of disorder, the transmission becomes impossible due to the compact localized states forming the flat band. 
When off-diagonal disorder is considered, the degeneracy of the band is preserved but the associated states are no longer confined to the unit cells. 
By perturbatively coupling the sender and receiver to the flat band, we derive a general effective Hamiltonian resembling a star network model with two hubs.
The effective couplings correspond to wavefunctions associated with the flat-band modes.
Specific relationships between these parameters define the quality of the quantum-state transfer which, in turn, are related to the degree of localization in the flat band. 
Our findings establish a framework for further studies of flat bands in the context of quantum communication. 
\end{abstract}

\maketitle

\section{Introduction}

Progress in 
quantum information processing have led us to the so-called noisy-intermediate-scale quantum era \cite{preskill18}. This means that the technology for assembling dozens of qubits to perform simple and proof-of-principle tasks is available \cite{arute19}. Yet, practical issues persist such as
faulty quantum gates and limited control over the processing units. 
A prominent
source of errors, besides decoherence, comes from the manufacturing process of quantum devices. For example, the parameters of a qubit network -- such as coupling strengths and local transition energies -- might deviate
from their original design, culminating in disorder.
This can consequently lead to Anderson localization of quantum information \cite{burrell07, allcock09}.
Since disorder cannot be fully suppressed, 
it is important to consider its influence on the quantum dynamics.
%

Quantum-state transfer (QST) and entanglement distribution are essential tasks to be performed on quantum networks
\cite{kimble08}.
A quantum communication channel can be set by a collection of spin-1/2 particles acting as qubits and linked via engineered exchange interactions. 
An arbitrary qubit state prepared at one end
of a 1D chain can be transmitted to the other end by the unitary evolution of the Hamiltonian. This idea was introduced by Bose in Ref. \cite{bose03} and many other schemes have since been proposed \cite{christandl04, plenio04, osborne04, wojcik05, wojcik07, li05, huo08, liu08, gualdi08, wang09,  banchi10, apollaro12, lorenzo13, paganelli13, lorenzo15, almeida16, almeida18, apollaro19}. 

A particular class of spin chains for QST relies on a complete engineering of their couplings. 
This approach results in a linear spectrum that supports end-to-end perfect state transfer 
in arrays of any size \cite{christandl04, plenio04}. 
High-fidelity QST protocols can also be designed 
with lower engineering costs by tuning the boundaries of a homogeneous spin chain \cite{wojcik05, wojcik07, banchi10}.
Other methods involve the application of strong local magnetic fields in order to effectively decouple the sender and receiver spins from the channel \cite{lorenzo13}. 

Under the influence of disorder, the overall performance 
of any QST scheme is expected to be reduced
to some degree. \cite{dechiara05,fitzsimons05,burgarth05,tsomokos07,giampaolo10,yao11,zwick12,zwick15, bruderer12, ashhab15, kay16,estarellas17}.
%
%
Chains featuring modified boundary couplings 
are more robust against static noise \cite{zwick12, almeida18}. When the communicating parties are perturbatively coupled to the channel, the conditions for 
an end-to-end effective interaction become more flexible because 
most of the channel modes barely interfere.
As such, it is possible to tune the end spins in a way that shields the dynamics from the influence of the strongly localized states \cite{almeida18pla,almeida19qinp}.
%
%
Alternatively, one can harness topological protection against disorder \cite{almeida16, estarellas17}.

In this paper, we go beyond the usual 1D schemes to explore a QST protocol on 
a diamond-like spin-1/2 chain [see Fig. \ref{fig1}] described by the $XX$ model.
Disordered quasi-1D materials, such as nanowires, have been shown to display peculiar strongly correlated phenomena \cite{petrovic16,gligori20, roy20}. 
Many quasi-1D networks are known to support flat bands \cite{hyrkas13,flach14,maimaiti17}. These are dispersionless Bloch bands
hosting macroscopic degeneracy, diverging density of states, zero-group velocity, and infinite effective mass \cite{derzhko15rev}.
A rich variety of transport regimes, including exotic Anderson transitions \cite{goda06,chalker10}, can emerge,
specially when the system is under the influence of perturbations that slightly lift the degeneracy \cite{souza09,vicencio13,leykam13,ramachandran17,khomeriki16,roy20,bouzerar21}. 

Here we consider a pair of communicating spins weakly coupled to a diamond channel that hosts a flat band.
In the ordered case, 
the band hosts a set of compact localized states, each restricted to one unit cell of the diamond lattice \cite{flach14,maimaiti17}. 
When off-diagonal disorder is present,
the flat band is preserved but 
is formed by a distinct set of eigenstates. These modes can be extended and thus mediate QST between the end cells.
Remarkably, we observe that
the competition between the (topological) compact localization and Anderson localization benefits the QST. Furthermore, an effective model is derived to explain that property. By solving it analytically, we highlight 
the key ingredients responsible for achieving better fidelities.
Our findings can be readily extended to other flat band bipartite lattices. 

%
%
 
\section{Model and flat-band structure}

\begin{figure}[t!] 
\includegraphics[width=0.42\textwidth]{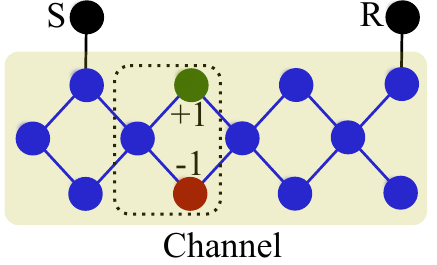}
\caption{\label{fig1} Sketch of the QST scheme. A quasi-1D diamond lattice functions as the channel.
Colored sites with alternating signs inside the unit cell (dashed square) depict the pattern of the compact localized states \cite{flach14,maimaiti17}. These occur in the ordered case and have the form $\ket{v_0^{(n)}} = \frac{1}{\sqrt{2}}(1,0,-1)$ for every cell $n$. 
Together, they compose a $N$-fold degenerate flat band at $E=0$.
Any amount of coupling (off-diagonal) disorder 
maintains the degeneracy given the bipartite nature of the diamond lattice. On the other hand, the spatial configuration of the flat-band modes will be modified due to their coupling with the dispersive modes. As a consequence, quantum-state transmission between spins $S$ and $R$ can be achieved.
}
\end{figure}

We consider a quantum channel that consists of $3N$ spin-1/2 particles arranged in a diamond-like configuration with open boundary conditions as shown in Fig. \ref{fig1}. They interact through a $XX$ Hamiltonian of the form ($\hbar = 1$) \begin{equation}
   H_{\mathrm{ch}} = \frac{1}{2}\sum_{\left< i, j \right>} J_{ij} (\hat{\sigma}_{i}^x\hat{\sigma}_{j}^x + \hat{\sigma}_{i}^y\hat{\sigma}_{j}^y) ,
\end{equation}
where $\hat{\sigma}_{i}^{x,y}$ 
are the usual Pauli operators for spin $i$ and
$J_{ij}$ is the nearest-neighbour interaction strength between spins $i$ and $j$. 
Herein, it is more convenient to visualize the channel as being composed of $N$ coupled vertical trimer cells. 
In the single-excitation sector, the Hamiltonian reads
\begin{align}
H_{\mathrm{ch}} &= \sum_{n=1}^{N} (J_{1,n}\ket{a_n}\bra{b_n}+J_{2,n}\ket{b_n}\bra{c_n}) \nonumber \\  
&\,\,\,\, +\sum_{n=1}^{N-1}(J'_{1,n}\ket{a_n}\bra{b_{n+1}}+J'_{2,n}\ket{c_n}\bra{b_{n+1}})+h.c.,
\label{Hch}
\end{align}
where $\ket{\ell_{n}}$ denotes a single spin flipped in the $n$-th cell at leg $\ell \in \lbrace a,b,c \rbrace$.

Let us now see how the flat band emerges.
Each cell contributes with the local eigenstates 
$\ket{v_0^{(n)}} = \frac{1}{\lambda_n}(J_{2,n},0,-J_{1,n})$ and $\ket{v_{\pm 1}^{(n)}} =\frac{1}{\lambda_n\sqrt{2}} (J_{1,n},\pm \lambda_n,J_{2,n})$, with corresponding eigenvalues $0$ and $\pm \lambda_n$, where $\lambda_n = \sqrt{J_{1,n}^2+J_{2,n}^2}$.
%
Within this basis set, transitions between different cells are given by:
\begin{align}
\bra{v_{\nu}^{(n+1)}}H_{\mathrm{ch}}\ket{v_{\pm 1}^{(n)}}&=\frac{\nu}{2\lambda_n}\left( J_{1,n}J'_{1,n}+J_{2,n}J'_{2,n} \right),\\
\bra{v_{\nu}^{(n+1)}}H_{\mathrm{ch}}\ket{v_{0}^{(n)}}&=\frac{\nu}{\sqrt{2}\lambda_n}\left( J_{2,n}J'_{1,n}-J_{1,n}J'_{2,n} \right),\label{transition}\\
\bra{v_{0}^{(n+1)}}H_{\mathrm{ch}}\ket{v_{0}^{(n)}}&=0,
\end{align}
where $\nu=\pm 1$. 

In the ordered case ($J_{i,n} = J'_{i,n} = J$) a quick look at the expressions above tells us that $\ket{v_0^{(n)}}$ are eigenstates of $H_{\mathrm{ch}}$ with the same energy $E=0$.
Indeed, they form a complete orthogonal basis at the center of the band. Given each $\ket{v_0^{(n)}}$ is spatially confined to the $n$-th unit cell, 
they are classified as compact localized states \cite{flach14,maimaiti17}. 
Therefore, a diamond channel with $N$ cells hosts a $N$-fold degenerate flat band at $E=0$.   


 
%
%

The ordered flat band cannot 
mediate a resonant QST \cite{wojcik07} between two external spins weakly coupled to the outermost cells (see Fig. \ref{fig1}).  
The compact localized states $\ket{v_{0}^{(n)}}$ forbid excitation transport between any pair of cells. 
This scenario changes, however, when $\ket{v_{0}^{(n)}}$ are no longer eigenstates of $H_{\mathrm{ch}}$. Any disorder in the channel will promote transitions between those and the dispersive modes $\ket{v_{\nu}^{(n)}}$ [Eq. (\ref{transition})]. 
In this work, instead of devising an engineering scheme for $H_{\mathrm{ch}}$ we will 
see how random fluctuations in the spin-spin couplings 
can assist a QST protocol.

Flat-band diamond lattices 
have been studied in the presence of diagonal as well of off-diagonal disorder \cite{leykam13, leykam17, roy20}. 
In the weak disorder regime, a general result is that the mixing between flat-band states and the others leads to a scaling of the localization length of the form $\xi \sim W^{-\gamma}$ at the flat band, with $W$ being the disorder width and the exponent $\gamma$ 
depending on the flat band class \cite{leykam13, leykam17}. 

One property
that deserves particular attention here is the bipartite nature of the diamond lattice. This means that we can
group the $N$ states $\lbrace \ket{b_n} \rbrace$ into one 
sublattice
and the remaining
$2N$ states $\lbrace \ket{a_n},\ket{c_n} \rbrace$ into another.
A known theorem \cite{sutherland86, inui94} states that bipartite lattices featuring only off-diagonal disorder sustains at least $M$ linearly independent states at $E=0$, where $M$ is the
difference between the number of sites 
of both sublattices.  
In addition, these $M$ states have no amplitude on the minority sublattice. For the diamond lattice, $M=N$.

The flat band is thereby preserved if we set
$J_{i,n}\rightarrow J_{i,n}(1+\delta_{i,n})$ and $J'_{i,n}\rightarrow J'_{i,n}(1+\delta'_{i,n})$, where  
$\delta_{i,n},\delta'_{i,n}$ are uncorrelated random numbers uniformly distributed in $[-W/2,W/2]$. However, note that while the 
$N$-fold degeneracy is maintained at $E=0$,
its corresponding modes are no longer
$\ket{v_0^{(n)}}$. Instead, we get another set of flat-band eigenstates 
$\ket{E_{FB,k}}$ which involves linear 
combinations of $\ket{v_0^{(n)}}$ and $\ket{v_\nu^{(n)}}$ but still 
have no amplitude on $\ket{b_n}$ \cite{sutherland86, inui94}.
Note that as our lattice is finite, there will always be one compact localized state left for the end cell, $\ket{v_0^{(N)}}$. Its contribution is negligible for our purposes. 
In the following section we show how the disordered flat band can mediate a QST protocol between the outermost cells.    

\section{Results}

\subsection{Effective Hamiltonian}

Let us now add two extra spins to the diamond channel, one at each end, to play the role of sender ($S$) and receiver ($R$). As shown in Fig. \ref{fig1}, they
are coupled to the sites $\ket{a_1}$ and $\ket{a_N}$, respectively. The full Hamiltonian of the system is now given by $H_{\mathrm{ch}}+H_{I}$, with the interaction Hamiltonian
\begin{equation}
H_I = g(\ket{S}\bra{a_1}+\ket{R}\bra{a_N}+ h.c.).
\end{equation}
Here we assume that $g$ is much smaller than the gap between the flat band and the 
non-zero energy states. 
The dispersion relation of the delocalized
states for an infinite ordered diamond lattice
reads $E(k)=\pm 2J\sqrt{1+\cos k}$, with $k$ being the typical wavenumber \cite{leykam13}. Then, the energy of the next non-zero energy states decreases as $\sim N^{-1}$ and so $g\ll J N^{-1}$. 
This ensures that only the flat-band modes will contribute to the QST dynamics. 

The perturbative coupling set by $g$ delivers a first-order resonant interaction involving $\ket{S}$, $\ket{E_{FB,k}}$ ($k=1,\ldots,N$), and $\ket{R}$. 
By generalizing the framework of single-mode resonant QST proposed in \cite{wojcik07}, we obtain the effective Hamiltonian
\begin{equation} \label{Heff}
H_{\mathrm{eff}} = g\sum_{k=1}^{N}(\mu_{1,k}\ket{S}\bra{E_{FB,k}}+\mu_{N,k}\ket{R}\bra{E_{FB,k}}+ h.c.),
\end{equation}
where the wavefunctions $\mu_{n,k} =\langle a_n \vert E_{FB,k} \rangle$ are now effective couplings.
Note that 
the states $\ket{c_1}$ and $\ket{c_N}$ can also host
the communicating parties without loss of generality. The states $\ket{b_n}$ have null amplitudes in the flat band and therefore are not suited for the QST protocol.

In matrix form, the Hamiltonian above reads
\begin{equation}\label{starnet}
H_{\mathrm{eff}}=
g\begin{pmatrix}
0 & 0 & \mu_{1,1} & \mu_{1,2} & \cdots & \mu_{1,N} \\
0 & 0 & \mu_{N,1} & \mu_{N,2} & \cdots & \mu_{N,N} \\
\mu_{1,1} & \mu_{N,1} & 0 & 0 &  & \vdots \\
\mu_{1,2} & \mu_{N,2} & 0 & 0 &   &  \\
\vdots & \vdots &  &  & \ddots &   \\
\mu_{1,N} & \mu_{N,N} & \cdots &  &  & 0
\end{pmatrix}.
\end{equation}
The effective QST model can be seen as a disordered star network with two hubs. 
We remark that the disorder featuring in the parameters
$\mu_{n,k}$ traces back to the fluctuations in $J_{i,n}$ and $J'_{i,n}$, with the constraint that $\eta_n = \sum_{k}|\mu_{n,k}|^2 \leq 1$.
%
The presence of disorder generally lead to 
$\eta_1\neq\eta_N$. We will see shortly that this population imbalance and the degree of localization of the flat band modes dictate the quality of the QST.
 

\subsection{Quantum-state transfer via flat bands}

We now analyze the transfer of an arbitrary qubit state $\ket{\psi} = \alpha \ket{0_S}+\beta\ket{1_S}$ prepared at site $S$ (see Fig. \ref{fig1}). The remaining spins are set in the ferromagnetic ground state such that $\ket{\Psi(t=0)}=\ket{\psi}\ket{0_{1}0_{2}\cdots 0_{3N}0_{R}}$. To evaluate the QST performance at site $R$, we compute the input-averaged (over all $\alpha$ and $\beta$) transfer fidelity \cite{bose03}:
\begin{equation}\label{avF}
F(t) = \frac{1}{2}+\frac{|f_{R}(t)|}{3}+\frac{|f_{R}(t)|^2}{6}.
\end{equation}
Hence, it suffices to track the evolution of the transition amplitude $f_{R}(t) = \langle R \vert U(t) \vert S \rangle$ over time, with $U(t) = e^{-iHt}$ being the quantum time evolution operator. 

We now turn our attention back to the effective two-hub star Hamiltonian [Eq. (\ref{starnet})] to obtain an expression for $f_{R}(t)$. Note that it embodies \textit{another} bipartite network featuring $N$ nodes in one sublattice (the flat band itself) and two nodes ($\ket{S}$ and $\ket{R}$) in the other. As such, 
by symmetry arguments \cite{sutherland86, inui94}, the spectrum 
is composed of the set of eigenvalues $\lbrace \pm\epsilon_1,\pm\epsilon_2, \lbrace 0 \rbrace_{N-2}  \rbrace$. The $N-2$ states at level $E=0$
do not have amplitude on the minority sublattice and thus will not contribute to the QST protocol.
The remaining four eigenstates can be written as
\begin{align}
\ket{\epsilon_{1}^{\pm}}&=\frac{1}{\sqrt{2}}\left( x_S\ket{S}+x_R\ket{R} \right) \pm \frac{1}{\sqrt{2}}\ket{\phi_1},\\
\ket{\epsilon_{2}^{\pm}}&=\frac{1}{\sqrt{2}}\left( y_S\ket{S}+y_R\ket{R} \right) \pm \frac{1}{\sqrt{2}}\ket{\phi_2},
\end{align}
where $\ket{\phi_i}$ are linear combinations of the states $\ket{E_{FB,k}}$, satisfying $\langle \phi_1 \vert \phi_2 \rangle = 0$. In each of those eigenstates 
the probability to find the excitation in either of the sublattices is 1/2. This is another remarkable symmetry property of bipartite networks \cite{souza17}. As $\lbrace \ket{\epsilon_{1}^{\pm}},\ket{\epsilon_{2}^{\pm}} \rbrace$ must be an orthonormal set, we also have that $y_S=-x_R^{*}$ and $y_R=x_S^{*}$, with $|x_R|^2+|x_S|^2=1$.

Now, expanding $U(t) = e^{-iHt}$ we obtain
\begin{align}
f_{R}(t)&=x_S^{*}x_R(\cos\epsilon_1 t-\cos\epsilon_2 t)\nonumber\\
&=-2x_S^{*}x_R \left[ \sin\left(\frac{\epsilon_1+\epsilon_2}{2}t\right)\sin\left(\frac{\epsilon_1-\epsilon_2}{2}t\right) \right].
\label{effFR}
\end{align}
The primary QST timescale will be dictated
by the slower sine function that depends on the gap $\delta \epsilon = \epsilon_1-\epsilon_2$.
The transition amplitude
$|f_{R}(t)|$ reaches its maximum at times 
$\tau=m\pi/\delta \epsilon$, where $m=n\delta\epsilon/(\epsilon_1 +\epsilon_2)$, with $n$ and $m$ being odd integers. (Strictly, $m$ is an integer only if  $\delta\epsilon/(\epsilon_1 +\epsilon_2)$ is rational.)
Given the QST time $\tau \propto g^{-1}$ and recalling that
$g\ll J N^{-1}$, in order to validate the effective Hamiltonian, then $\tau$ is typically larger than $O(N)$.

\begin{figure}[t!] 
\includegraphics[width=0.4\textwidth]{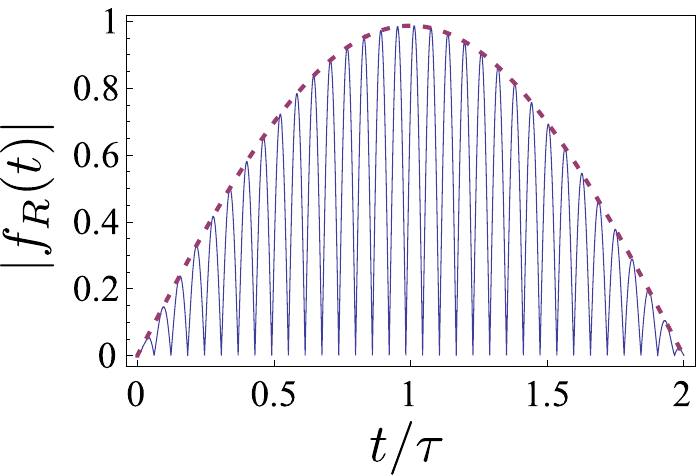}
\caption{\label{fig2} Time evolution of the transition amplitude $|f_R(t)|$. 
The solid curve is obtained by exact numerical diagonalization of the full Hamiltonian $H_{\mathrm{ch}}+H_{I}$ for $N=10$ cells, $g=0.01J$, and $W=0.2J$. Only one (typical) disorder realization is shown. The dashed curve represents 
$C_{S,R}|\sin (\delta \epsilon t/2)|$, which is the slow oscillating part of the analytical expression for $f_R(t)$ in Eq. (\ref{effFR}). Both $C_{S,R}$ and $\delta\epsilon$ are obtained numerically.
Time is expressed in units of the QST time $\tau = \pi/\delta \epsilon$.
}
\end{figure}

The maximum amplitude that $f_{R}(t)$ can achieve is related to a correlation parameter $C_{S,R}=2|x_S^{*}x_R|=4|\langle \epsilon_{i}^{\pm} \ket{S}\bra{R} \epsilon_{i}^{\pm} \rangle|$, which ranges from 0 to 1. It can thus be used to assess the quality of the QST.
Figure \ref{fig2} depicts the time evolution 
of the transition amplitude during a QST cycle as obtained by exact diagonalization of the full Hamiltonian for a single disorder sample. The wave envelope given by $C_{S,R}|\sin (\delta \epsilon t/2)|$
is also plotted for comparison.

A perfect QST [within the effective framework of Eq. (\ref{starnet})], with $F(\tau)=1$ [Eq. (\ref{avF})], can only be achieved provided $C_{S,R}=1$, which implies $|x_{S}|=|x_{R}|=1/\sqrt{2}$.
With regard to the effective couplings $\mu_{n,k}$, a particular condition must be fulfilled. 
To see this, we can solve the eigenvalue equation $H_{\mathrm{eff}}\ket{\epsilon_{i}^{\pm}}=\pm\epsilon_{i}\ket{\epsilon_{i}^{\pm}}$ analytically to obtain:
\begin{align}
\epsilon_{1,2}&=g\left[\frac{1}{2}\left(A\pm\sqrt{A^2-4B}\right)\right]^{1/2},\\
x_{S}&=\left( 1-\frac{\Lambda^2}{(\tilde{\epsilon_i}^2-\eta_N)^2+\Lambda^2} \right)^{1/2},\\
x_{R}&=\frac{\sqrt{2}\Lambda}{\tilde{\epsilon_i}^2-\eta_N} x_{S},
\end{align}
where $A=\eta_1+\eta_N$, $B=\eta_1\eta_N-\Lambda^2$, $\Lambda=\sum_{k}(\mu_{1,k}\mu_{N,k})$, and $\tilde{\epsilon_{i}}=\epsilon_{i}/g$. Hence, the correlation parameter can be written as
$C_{S,R}=2/\sqrt{4+\Delta^2}$, where $\Delta=(\eta_1-\eta_N)/\Lambda$. We immediately see that $C_{S,R}=1$ whenever $\eta_1=\eta_N$ (as long as $\Lambda \neq 0$). 
In such a case, the fluctuations in the parameters $\mu_{n,k}$ are irrelevant. 
This is remarkable from the standpoint of the effective Hamiltonian in Eq. (\ref{starnet}). It means that 
in principle one can realize an almost perfect QST despite any level 
of disorder in $\mu_{n,k}$ by tuning a \textit{single} parameter.

\begin{figure}[t!] 
\includegraphics[width=0.35\textwidth]{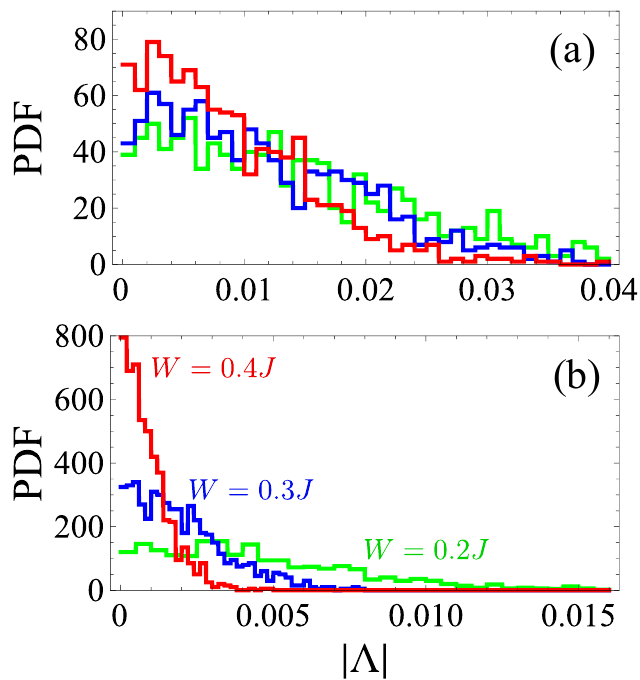}
\caption{\label{fig3} 
Probability density functions of the parameter 
$|\Lambda|=|\sum_{k}(\mu_{1,k}\mu_{N,k})|$ evaluated numerically via exact diagonalization of $H_{\mathrm{ch}}$
considering (a) $N=20$ and (b) $N=40$ cells
for $10^3$ independent disorder samples. The disorder widths $W$ considered in both panels are $0.2J$, $0.3J$, and $0.4J$. 
}
\end{figure}

Here, we cannot manipulate $\mu_{n,k}$ directly, though. These parameters are attached to the flat-band modes of the physical lattice. Our goal now is to harness the randomness present in the exchange couplings $J_{i,n}$ and $J'_{i,n}$. 
By doing so, we generally obtain $\eta_1 \neq \eta_N$ and then
$\Delta \rightarrow 0$ is required to attain
$C_{S,R}\rightarrow 1$. 
Note that $\Delta$ is inversely proportional to the parameter 
$|\Lambda|=|\sum_{k}(\mu_{1,k}\mu_{N,k})|$. The latter scans 
the whole flat band modes via their amplitudes on states $\ket{a_1}$ and $\ket{a_N}$ (the ones to which spins $S$ and $R$ are coupled). In a disordered system, we expect that the product of two wavefunctions such as $\mu_{1,k}\mu_{N,k}$ be typically close to zero, especially for spins
residing at distant locations.

In Fig. \ref{fig3} we show probability density functions (PDFs) 
of $|\Lambda|$ for $N=20,40$ cells and selected values of the disorder width $W$. Indeed,
the distributions become more peaked around zero as $W$ grows. This trend is more severe for 
larger system sizes [Fig. \ref{fig3}(b)]. Despite
the correlation $C_{S,R}$ is a function of the ratio $\Delta=(\eta_1-\eta_N)/\Lambda$, the factor $\Lambda$ is the one that counts as far as $N$ is concerned. This can be observed from the fact that $\eta_n$ is a local quantity.   

\begin{figure}[t!] 
\includegraphics[width=0.4\textwidth]{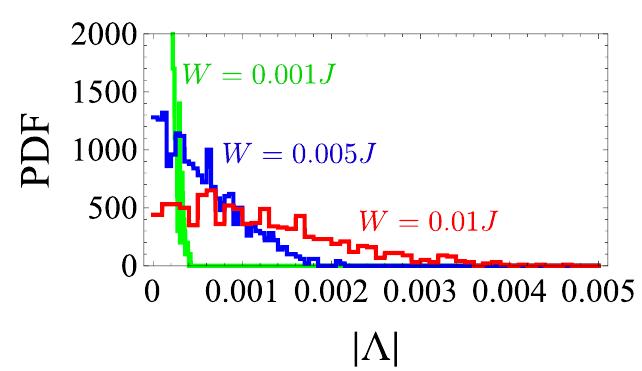}
\caption{\label{fig4} 
Probability density functions of 
$|\Lambda|=|\sum_{k}(\mu_{1,k}\mu_{N,k})|$ for the weak disorder regime. 
Each distribution is obtained for
$10^3$ independent disorder samples considering $N=20$. The disorder widths $W$ are indicated by each curve. 
}
\end{figure}

Things are different, however, as $W\rightarrow 0$. In this regime, we observe the opposite behavior for $\Lambda$. The increasing of $W$ actually benefits the formation of $C_{S,R}$.
The PDF for very low disorder widths is shown in Fig. \ref{fig4}. Such a reverse trend can be understood as a reminiscent influence of the compact localized states, to which $\Lambda=0$, as we depart from $W=0$. 
As the disorder increases, the flat-band modes become strongly localized again, but are no longer restricted to each cell. Instead, their localization length scales as $\xi \sim W^{-\gamma}$ \cite{leykam13,leykam17}.
It is between those two regimes that the QST will occur.


\subsection{Protocol performance}

\begin{figure}[t!] 
\includegraphics[width=0.35\textwidth]{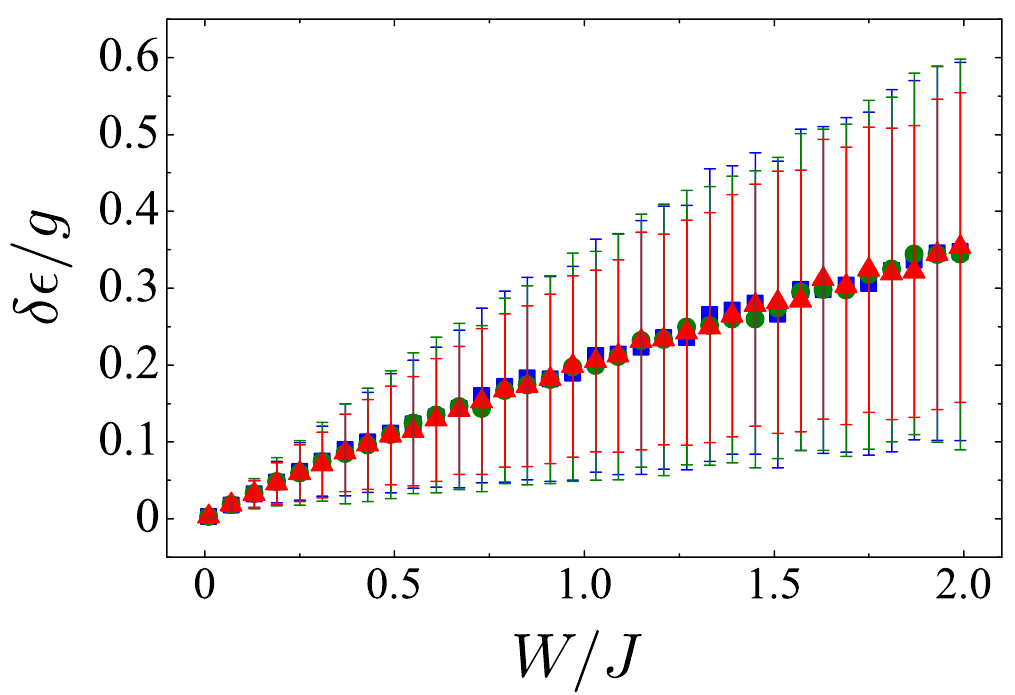}
\caption{\label{fig5} Energy gap $\delta \epsilon / g=(\epsilon_1-\epsilon_2)/g$ versus the disorder width $W$ for 
$N=10$ (blue squares), 
$N=20$ (green circles), and $N=40$ (red triangles). Data are obtained via exact numerical diagonalization of the full Hamiltonian $H_{\mathrm{ch}}+H_I$, considering $g=0.01J$, and averaged over $10^{3}$ independent realizations of disorder.
Error bars are the corresponding mean absolute deviations.
}
\end{figure}

Now that all the relevant quantities that govern the speed and quality of the QST have been presented, 
we are ready to  
the test its performance against $W$.

Any disorder induces fluctuations in the energy spectrum and affects the transfer time $\tau \propto \delta\epsilon^{-1}$. In Fig. \ref{fig5} we show
how the gap $\delta\epsilon = \epsilon_1-\epsilon_2$ (in units of $g$) responds to $W$. It grows roughly linear with $W$, not being affected by the number of cells $N$. 
Another caveat to the limit $W\rightarrow 0$ is that a vanishing gap 
implies in a extremely slow QST, which is not a desirable feature. Therefore, disorder is needed to bypass the compact localized states of the flat band and also 
to make the QST faster. 

Because the fluctuations in the gap increase with $W$, we track the QST fidelity over a given time window (instead of a specific time).  
 Let us define $F_{\mathrm{max}} = \mathrm{max}\lbrace F(t) \rbrace$ as the maximum fidelity achieved for $t \in [0,t_{\mathrm{max}}]$, where $t_{\mathrm{max}}=20\pi/g$. Given $\tau = m \pi/\delta \epsilon$, we remark that
 the value of $t_{\mathrm{max}}$ corresponds to $m=1$ and $\delta\epsilon / g =0.05$. In this way, 
 the QST fidelity for disorder widths slightly below $W\approx 0.2J$ (cf. Fig. \ref{fig5}) may be underestimated. 
 For $W>0.2J$, $t_{\mathrm{max}}$ is enough for the first QST cycle to occur [see Eq. (\ref{effFR})].  

\begin{figure}[t!] 
\includegraphics[width=0.4\textwidth]{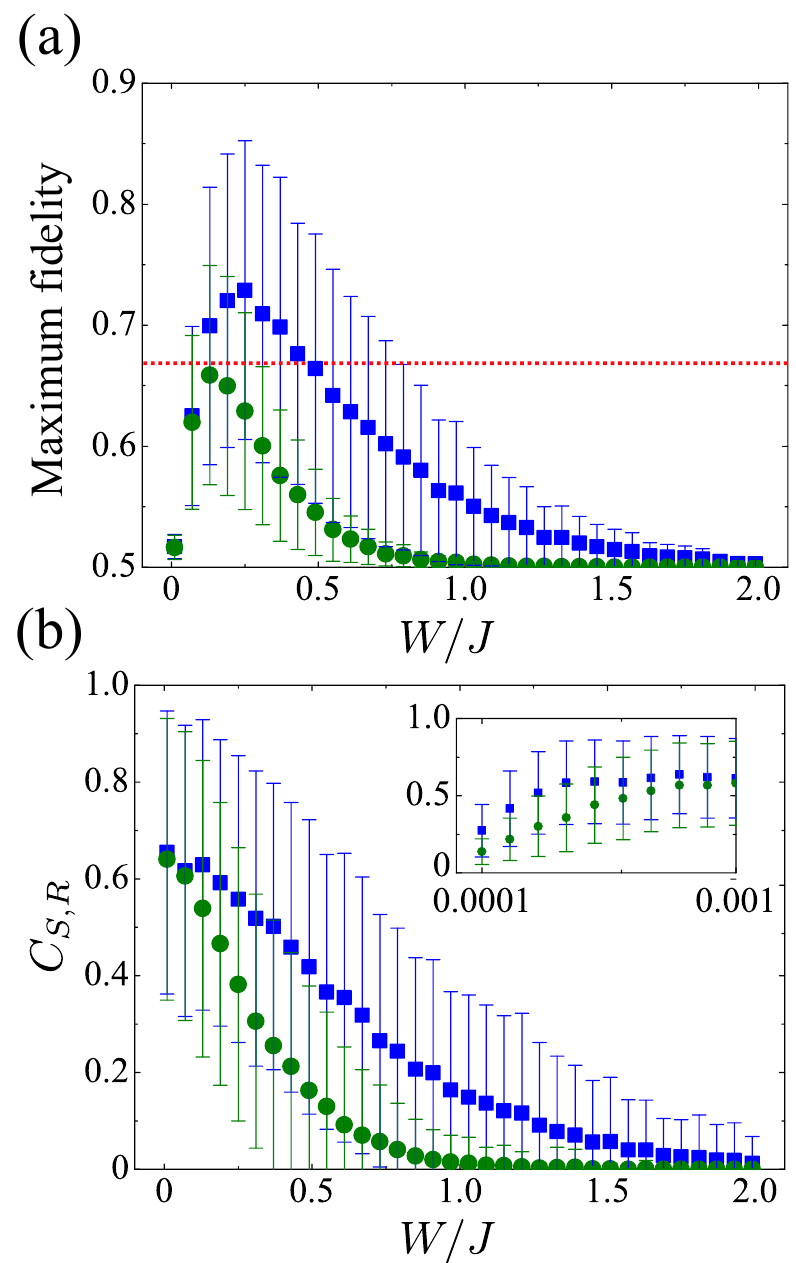}
\caption{\label{fig6} (a) Maximum fidelity 
$F_{\mathrm{max}} = \mathrm{max}\lbrace F(t) \rbrace$, evaluated for $t \in [0,20\pi/g]$,
against $W$. Blue squares (green circles) denote the data for $N=10$ ($N=20$), obtained via exact numerical diagonalization of the full Hamiltonian with $g=0.01J$. The dotted horizontal bar indicates the fidelity threshold associated to classical transmission, $F=2/3$. 
(b) Correlation between spins $S$ and $R$, $C_{S,R}$, also obtained for $g=0.01J$. The inset shows the same measure for smaller values of $W$. (The first point lies at $W=0.01J$ in the main figures.)    
Both quantities are averaged over $10^{3}$ independent realizations of disorder. Vertical error bars represent the mean absolute deviations.  
}
\end{figure}

\begin{figure}[t!] 
\includegraphics[width=0.4\textwidth]{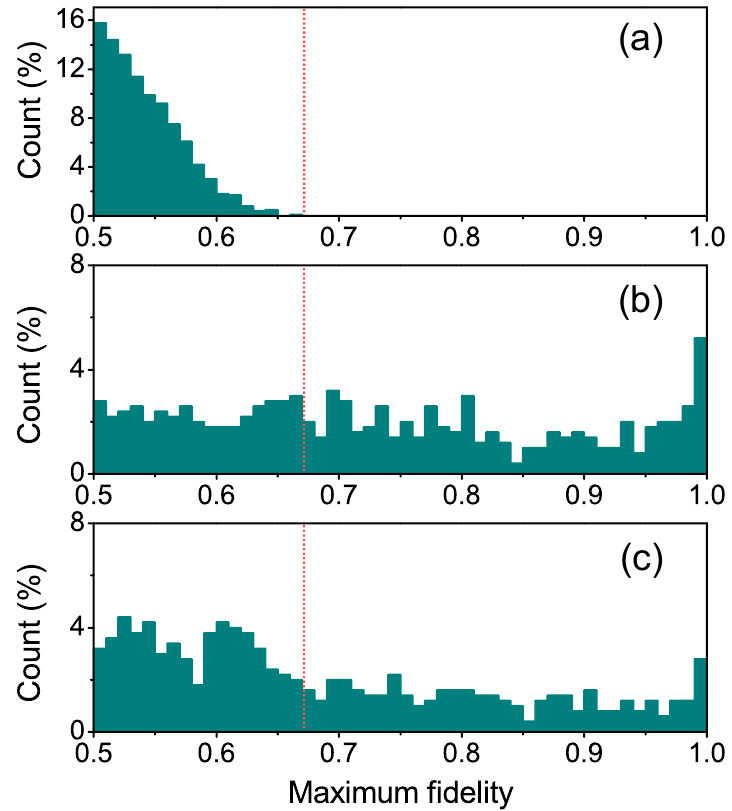}
\caption{\label{fig7} Histograms of 
the maximum QST fidelities based on 
$10^{3}$ independent realizations of disorder for $g=0.01 J$, $N=10$ cells, (a) $W=0.01J$, (b) $W=0.2J$, and (c) $W=0.4J$. Dotted vertical bars represent the fidelity threshold for classical transmission, $F=2/3$.
}
\end{figure}

The results for the QST fidelity are displayed in Fig. \ref{fig6}(a) considering $N=10$ and $N=20$ cells and fixed $g=0.01J$. 
Note that this value of $g$ is 
compatible with the system sizes considered.  
All the elements discussed so far are manifested 
through the fidelity performances. Indeed, 
the overall QST quality declines in 
the larger system size. This have been predicted 
in the analysis of the parameter $\Lambda$ (Fig. \ref{fig3}).
Yet, it is possible to reach fidelities above the classical threshold of $2/3$ \cite{bose03,horodecki99} at intermediate disorder levels.
Figure \ref{fig7} 
displays histograms of the maximum fidelities for $N=10$ and some selected values of $W$.

We remark that the 
poor performances associated to the 
lower values of $W$ in Fig. \ref{fig6}(a)
is a consequence of the chosen time interval. 
To confirm this, Fig. \ref{fig6}(b) shows the behavior of correlation $C_{S,R}$ against $W$. As we have seen, $C_{S,R}$ ultimately determines the quality of the QST.
Therefore, in this case higher fidelities \textit{can} be achieved at times $t>t_{\mathrm{max}}$. But if we were to consider disorder levels $W\ll 0.001J$ [inset of Fig. \ref{fig6}(b)], then
the QST would be unfeasible because of the reverse localization trend discussed earlier (see Fig \ref{fig4}).  
Indeed, $C_{S,R}$ must vanish as $W\rightarrow 0$ so as to conform with the development of the compact localized states. 


\section{Concluding remarks}

We studied a resonant QST through a flat band hosted by a disordered diamond lattice. In particular, off-diagonal disorder was considered 
for preserving the flat band due to the bipartite topology of the lattice.
Our findings revealed that the QST protocol
yields
good fidelities when a certain amount of disorder is present. 
The underlying phenomenon is a transition
of the flat-band modes from compact localization, when $W=0$, to Anderson localization as $W$ increases \cite{leykam13,leykam17}. For intermediate levels of disorder, those modes can 
jointly sustain significant amplitudes on distant cells.
 While we considered a simple uniform distribution for the disorder, similar results are obtained for uncorrelated Gaussian-distributed disorder (which is a more realistic situation).  
 
We also derived and solved an effective Hamiltonian [Eq. (\ref{starnet})] that applies for a pair of sites perturbatively coupled to any flat-band bipartite lattice. The model resembles a star network with two hubs and disordered couplings associated to the flat-band wavefunctions.
By deriving analytical expressions for the relevant eigenstates, we were able to 
identify the parameters that control the QST. 
Interestingly, if $\eta_1 = \eta_N$ then an almost perfect QST can occur for very small $g$ despite the level of fluctuations associated to the effective couplings $\mu_{n,k}$. 

The effective model we addressed is a powerful tool to study quantum transport via flat bands.  
It may be explored on its own as a synthetic network aiming for the remarkable relationship between the couplings. 
%
We hope that our results encourage further research on quantum communication in other classes of flat bands \cite{flach14,leykam17}.


\section{Acknowledgments}

This work is supported by CNPq,
CAPES, FINEP, CNPq-Rede Nanobioestruturas, 
and FAPEAL. 


%

\end{document}